\documentclass[aps,amsmath,twocolumn,notitlepage,amssymb,prb,longbibliography]{revtex4-1}

\usepackage{graphicx}% Include figure files
\usepackage{dcolumn}% Align table columns on decimal point
\usepackage{bm}% bold math
\usepackage[bookmarks=true,colorlinks,linkcolor=blue,urlcolor=blue,citecolor=red]{hyperref}
\usepackage{wasysym}
\usepackage{bbold}
\usepackage{stmaryrd}
\usepackage{verbatim}
\usepackage{subfigure}

\newcommand{\bs}{\boldsymbol}

\newcommand{\p}{\prime}

\newcolumntype{C}[1]{>{\centering\let\newline\\\arraybackslash\hspace{0pt}}m{#1}}

\begin{document}
\title{Detecting spin fractionalization in a spinon Fermi surface spin liquid}
\author{Yao-Dong Li$^{1}$}
\author{Gang Chen$^{1,2}$}
\email{gangchen.physics@gmail.com}
\affiliation{$^{1}$State Key Laboratory of Surface Physics, 
Department of Physics, 
Center for Field Theory \& Particle Physics,
Fudan University, Shanghai, 200433, China}
\affiliation{$^{2}$Collaborative Innovation Center of Advanced Microstructures,
Nanjing, 210093, China}

\date{\today}

\begin{abstract}
Motivated by the recent proposal of the spinon Fermi surface spin liquids for 
several candidate materials such as YbMgGaO$_4$, we explore the experimental 
consequences of the external magnetic fields on this exotic state. Specifically, 
we focus on the weak field regime where the spin liquid state is well preserved 
and the spinon remain to be a good description of the magnetic excitations. 
From the spin-1/2 nature of the spinon excitation, we predict the unique features 
of the spinon continuum when the weak magnetic field is applied to the system.
Due to the small energy scale of the exchange interactions between the local 
moments in the spin liquid candidate like YbMgGaO$_4$, our proposal for the 
spectral weight shifts and spectral crossing in the magnetic fields can be 
immediately tested by inelastic neutron scattering experiments. Several other 
experimental aspects about the spinon Fermi surface and the spinon excitations 
are discussed and proposed. Our work provides an experimental scheme to examine 
the fractionalized spinon excitation and the candidate spin liquid states in 
YbMgGaO$_4$, the 6H-B phase of Ba$_3$NiSb$_2$O$_9$ and other relevant materials.
\end{abstract}

\maketitle

\section{Introduction}
\label{sec1}

A quantum spin liquid (QSL) is an exotic quantum phase of
matter that carries long-range quantum entanglements in 
the modern terms. The association with 
the long-range quantum entanglements
has not yet led to the observable effects for QSLs. 
A more common description of the QSLs often involves certain 
emergent gauge structure and the fractionalized spinon 
excitation~\cite{PALee,Balents,savary2016quantum}.   
The experimental search of QSLs has lasted for forty years  
since the original proposal by Anderson in 
1973~\cite{Anderson1973,Anderson1987}. 
The absence of magnetic order is often used as 
the first diagnosis of QSLs in experiments. 
Many QSL candidate materials have been proposed so far, 
but the direct confirmation of QSLs has not been 
achieved in any of these materials. More recently, 
several materials including YbMgGaO$_4$ and 6H-B phase of Ba$_3$NiSb$_2$O$_9$, 
have been proposed to realize a spinon Fermi surface spin liquid~\cite{YaoShenNature,YueshengmuSR,Yaodong201612,PhysRevB.95.060402,PhysRevB.84.180403}, 
an exotic state that was originally proposed for the triangular 
lattice organic materials ${\kappa}$-{(ET)}$_{2}${Cu}$_{2}$(CN)$_{3}$ 
and EtMe$_3$Sb[Pd(dmit)$_2$]$_2$~\cite{organictherm,organics2,kappaET,dmit,LeeLeePRL,Motrunich2005}.

The well-known QSL candidates ${\kappa}$-{(ET)}$_{2}${Cu}$_{2}$(CN)$_{3}$ 
and EtMe$_3$Sb[Pd(dmit)$_2$]$_2$ are proximate to the Mott transition from 
the superconductor or metal state to the Mott insulating 
state~\cite{organictherm,organics2,kappaET,dmit,PhysRevLett.95.177001,Furukawa1707}. 
Due to the proximity to the Mott transition, the charge gap is small 
and the charge fluctuation is strong, which induces a sizable four-spin  
ring exchange interaction~\cite{Motrunich2005}. This ring exchange 
interaction competes the pair-wise Heisenberg interaction and frustrates 
the 120-degree magnetic order. These competing interactions were suggested
to be the driving force to enhance the quantum fluctuation of the spin 
degrees of freedom and help stabilize the QSL ground state in the 
organic materials~\cite{LeeLeePRL,Motrunich2005,PhysRevB.73.155115}.   
Based on the proximity to the Mott transition, the spinon Fermi surface
U(1) QSL and their instabilities~\cite{LeeLeePRL,Motrunich2005,PhysRevB.73.155115,PhysRevLett.98.067006,PhysRevB.76.165104,PhysRevB.87.045119,PhysRevB.83.235119,PhysRevLett.106.056402,PhysRevB.81.245121,PhysRevLett.104.066403,PhysRevB.91.115111,PhysRevB.78.045109,PhysRevB.84.165126,PhysRevB.94.115113} were suggested to provide a resonable and compelling description 
and/or theoretical prediction of various experimental results such as the 
thermodynamic and thermal transport properties~\cite{ItouNaturePhys,Yamashita2011,organictherm,organics2,kappaET,dmit,PhysRevLett.95.177001,Furukawa1707}. 
Despite being promising QSL candidates, these organic materials have 
a rather low spin concentration that prevents the data-rich inelastic 
neutron scattering measurement at the current laboratory setting. 
Thus, the full momentum and energy resolved spectroscopic information 
of the magnetic excitations in these materials is lacking.

In contrast to the weak Mott insulating phase of the organics, 
both the rare-earth triangular lattice antiferromagnet 
YbMgGaO$_4$ and the spin-1 antiferromagnet 6H-B Ba$_3$NiSb$_2$O$_9$ are in     
the strong Mott regime where the charge fluctuation is weak. 
The underlying QSL physics, if there is, should be fundamentally
different from the organics. Let us begin with YbMgGaO$_4$. 
This material was first discovered in the powder form and 
proposed as a gapless QSL~\cite{srep}, 
and was later suggested as the first QSL candidate in the 
{\it strong spin-orbit-coupled Mott insulator} with odd electron 
fillings~\cite{Yuesheng2015,YaodongPRB,YaoShenNature,Yaodong201608,Yaodong201612}.
The later proposal is compatible with the more fundamental 
view based on the time reversal symmetry and quantum 
entanglements~\cite{PNAS,Yuesheng2015,YaodongPRB,YaoShenNature,Yaodong201612}
that states the ground state of strong spin-orbit-coupled Mott insulator
with odd electron fillings must be exotic provided the presence of time
reversal symmetry. Here, the Yb local moments in YbMgGaO$_4$ remain disordered 
down to the lowest measured temperature at which the magnetic entropy is almost 
exhausted~\cite{srep,YueshengmuSR,YaoShenNature,Martin2016,Shiyan2016}. 
The low-temperature heat capacity with ${C_v \propto T^{2/3}}$ 
and the dispersion of the dynamic spin structure from the inelastic 
neutron scattering on single-crystal samples found a reasonable 
agreement with the theoretical prediction for the spinon Fermi surface 
state~\cite{YaoShenNature,Yaodong201612,Patrick1992,LeeLeePRL,Motrunich2005}.
Unlike YbMgGaO$_4$, the 6H-B phase of Ba$_3$NiSb$_2$O$_9$ has 
spin-1 local moments~\cite{PhysRevLett.107.197204,PhysRevB.93.214432,PhysRevB.95.060402,PhysRevLett.109.016402,PhysRevB.86.224409,PhysRevB.86.104411,PhysRevLett.108.087204,PhysRevB.84.180403}. 
Nevertheless, both earlier theoretical suggestion and the recent inelastic neutron scattering 
experiments propose that the spinon Fermi surface QSL is realized in this material~\cite{PhysRevB.84.180403,PhysRevB.95.060402}. 

In general, there are two major questions concerning the candidate QSL
ground states of YbMgGaO$_4$ and 6H-B Ba$_3$NiSb$_2$O$_9$. The first 
and probably the most crucial one is whether the excitation continuum 
in the inelastic neutron scattering is truly a spinon continuum and 
represents the spin quantum number fractionalization. The second question 
is the microscopic mechanism for the QSL behavior in these materials. 
For YbMgGaO$_4$, it was suggested that the anisotropic interaction 
of the local moments, due to the spin-orbit entanglement, could 
enhance the quantum fluctuation and destabilize the 
ordered phases~\cite{Yuesheng2015,YaodongPRB,Yaodong201608}. 
These two questions have been partially addressed by the mean-field 
theory~\cite{YaoShenNature} and the later projective symmetry group
analysis~\cite{YaoShenNature,Yaodong201612} that identify the
spinon Fermi surface U(1) QSL as the candidate ground state
for YbMgGaO$_4$. For 6H-B Ba$_3$NiSb$_2$O$_9$, 
the phenomenological approach based on fermionic partons with competing 
exchange interactions also suggests a spinon 
Fermi surface state~\cite{PhysRevB.95.060402,PhysRevB.84.180403}.

Ideally, it would be nice to directly solve the relevant microscopic spin 
model and see if one can obtain any QSL ground state in the phase diagram,
then both questions may be completely resolved. Due to the complication of
the models, this is difficult 
even numerically~\cite{YaodongPRB,Yaodong201612,1703,PhysRevB.95.165110}.
In this work, instead of directly tackling the microscopic spin 
model~\cite{YaodongPRB,Yaodong201608,Liu2016}, we work on the spinon 
mean-field Hamiltonian~\cite{YaoShenNature,Yaodong201612} that
has provided a reasonable description of the inelastic neutron 
scattering results for YbMgGaO$_4$. To ensure the nature of the spinon 
continuum in the inelastic neutron scattering results,
we propose a simple experimental scheme to test the spin quantum 
number fractionalization and confirm the spinon excitations. 
We suggest to apply a {\it weak external magnetic field} and study 
the spectral weight shifts of the dynamic spin structure factor. 
The splitting of the degenerate spinon bands by the magnetic 
field is directly revealed by the spinon particle-hole continuum 
that is detected by the dynamic spin structure factor. 
We show that the persistence of the spinon continuum,
the spectral weight shifts and the spectral crossing around the 
$\Gamma$ point, the existence of the upper and lower excitation edges
under the weak magnetic field represent unique 
properties of the spinon excitation for the spinon Fermi surface
state, and thus provide a visible experimental prediction for the 
identification of the spinon excitation with respect to 
the spinon Fermi surface. In the bulk of the paper, we illustrate 
this idea with an effective phenomenological spinon model for YbMgGaO$_4$, 
and the approach can be adjusted to other systems with little modification. 
 
The remaining part of the paper is organized as follows. 
In Sec.~\ref{sec2}, we explain our view on the magnetic excitation continuum and 
the weak spectral weight in the inelastic neutron scattering results on 
YbMgGaO$_4$ and motivate our approach in this paper. In Sec.~\ref{sec3}, 
we justify the mean-field Hamiltonian in the magnetic field.
In Sec.~\ref{sec4}, we obtain the dynamic spin structure factor from 
the free spinon theory in the magnetic field and explain the spectral 
weight shifts. In Sec.~\ref{sec5}, we include the spinon interactions
with a random phase approximation in our results. Finally in Sec.~\ref{sec6}, 
we conclude with a discussion about various future experimental direction 
for the spinon Fermi surface state.

\section{The spinon Fermi surface state}
\label{sec2}

We start with the fermionic parton representation for the spin operator such
that
\begin{equation}
{\boldsymbol S}_i^{} = \frac{1}{2} f^\dagger_{i\alpha}
{\boldsymbol{\sigma}}^{}_{\alpha\beta} f^{\phantom\dagger}_{i\beta} ,
\end{equation}
where $f^{\dagger}_{i\alpha}$ ($f^{\phantom\dagger}_{i\alpha}$)
creates (annihilates) one spinon with spin $\alpha$ ($=\uparrow,\downarrow$)
at the lattice site $i$ and ${\boldsymbol{\sigma} = (\sigma^x,\sigma^y,\sigma^z)}$
is a vector of Pauli matrices. This construction is further supplemented by
a Hilbert space constraint ${\sum_{\alpha} f^\dagger_{i\alpha}
f^{\phantom\dagger}_{i\alpha} = 1}$. At the mean-field level,
the following spinon Hamiltonian,
\begin{eqnarray}
H_{\text{MF}}^{} &=& - t_1 \sum_{\langle ij \rangle,\alpha}
f^{\dagger}_{i\alpha} f^{}_{j\alpha}
- t_2 \sum_{\langle\langle ij \rangle\rangle,\alpha}
f^{\dagger}_{i\alpha} f^{}_{j\alpha}
\nonumber \\
&& - \mu \sum_{i,\alpha} f^{\dagger}_{i\alpha} f^{}_{i\alpha}
\label{Hmf}
\end{eqnarray}
was proposed for YbMgGaO$_4$ and gives a large spinon Fermi 
surface~\cite{YaoShenNature,Yaodong201612}. Here we introduce 
the second neighbor spinon hopping in addition to the first 
neighbor spinon hopping on the lattice, although the first
neighbor spinon hopping is sufficient to give the spinon 
Fermi surface state. Adding the second neighbor spinon hopping
could give a mean field state with a better variational energy. 
Moreover, the spin rotational symmetry
for the first and the second neighbor spinon hoppings in Eq.~\eqref{Hmf}
is required by the lattice symmetry that is realized projectively
for the spinons~\cite{Yaodong201612}.
Finally, the chemical potential $\mu$ is introduced to impose the
Hilbert space constraint.

It was found that the spinon particle-hole 
excitation of the simple spinon Fermi surface ground state
of Eq.~\eqref{Hmf} provides a consistent magnetic 
excitation continuum with the inelastic neutron scattering 
experiments. Moreover, the anisotropic spin interaction, that is 
included into the the spinon mean-field theory by a random phase 
approximation (RPA), gives a weak spectral peak at the M points, 
which is consistent with the experimental 
observation~\cite{YaoShenNature,Martin2016}.

The spinon continuum itself is certainly more important than 
the weak spectral peak at low energies. The weak spectral peak 
at certain momenta merely represents some collective mode of 
the spinons that is enhanced by the residual and short-range 
interaction between the fermionic spinons, and is quite common 
for example in the Fermi liquids of electrons as an analogy. 
Nevertheless, the spectral peak does provide hints about the 
form of the microscopic interactions. In contrast, the spinon 
continuum is a consequence of the spin quantum number 
fractionalization that reveals the defining nature of QSLs.

Since we think the spinon continuum is more important and
the spinon continuum is already obtained by the free-spinon
theory of $H_{\text{MF}}^{}$, our approach will mostly rely
on the free-spinon mean-field theory and focus on the
spinon continuum rather than the weak spectral peak. 
The (short-range) anisotropic spin interaction will be included 
into the free-spinon theory in the later parts of the paper.
The coupling to the gapless U(1) gauge photon is not included 
throughout this paper. Nevertheless, the spinon-gauge coupling 
should have an important effect on the low-energy properties 
of the system~\cite{Patrick1992,LeeLeePRL}. 

%-----------------------------
\begin{figure}[t]
{
\includegraphics[width=4.2cm]{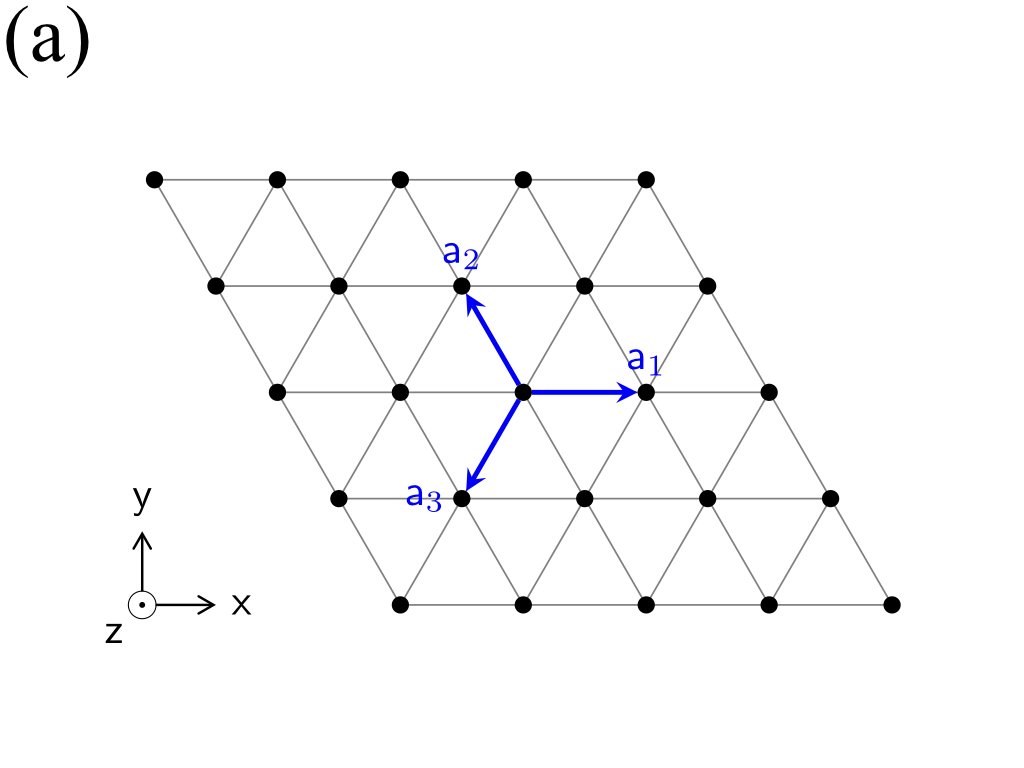}
\includegraphics[width=4.2cm]{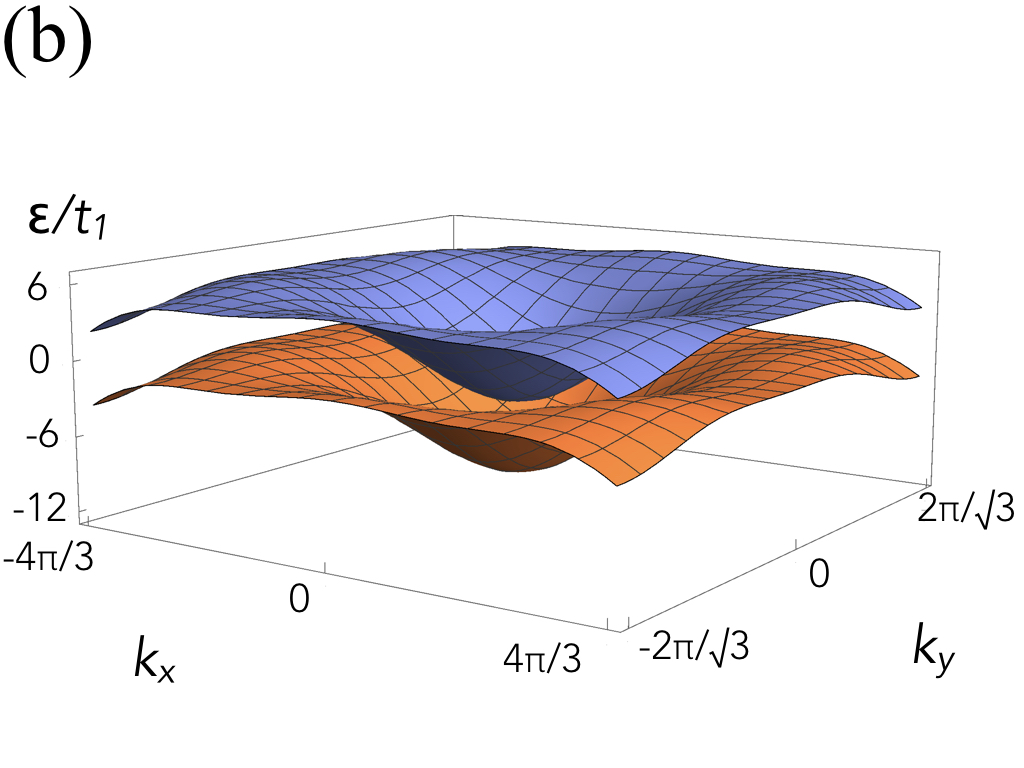}
\includegraphics[width=4.2cm]{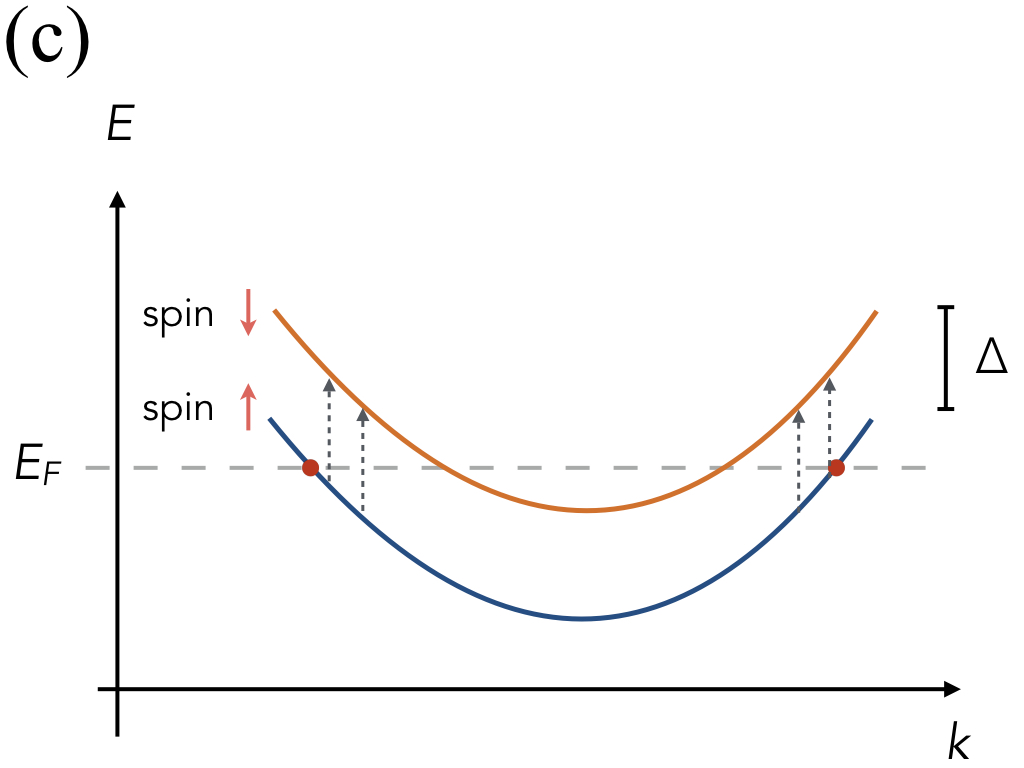}
\includegraphics[width=4.2cm]{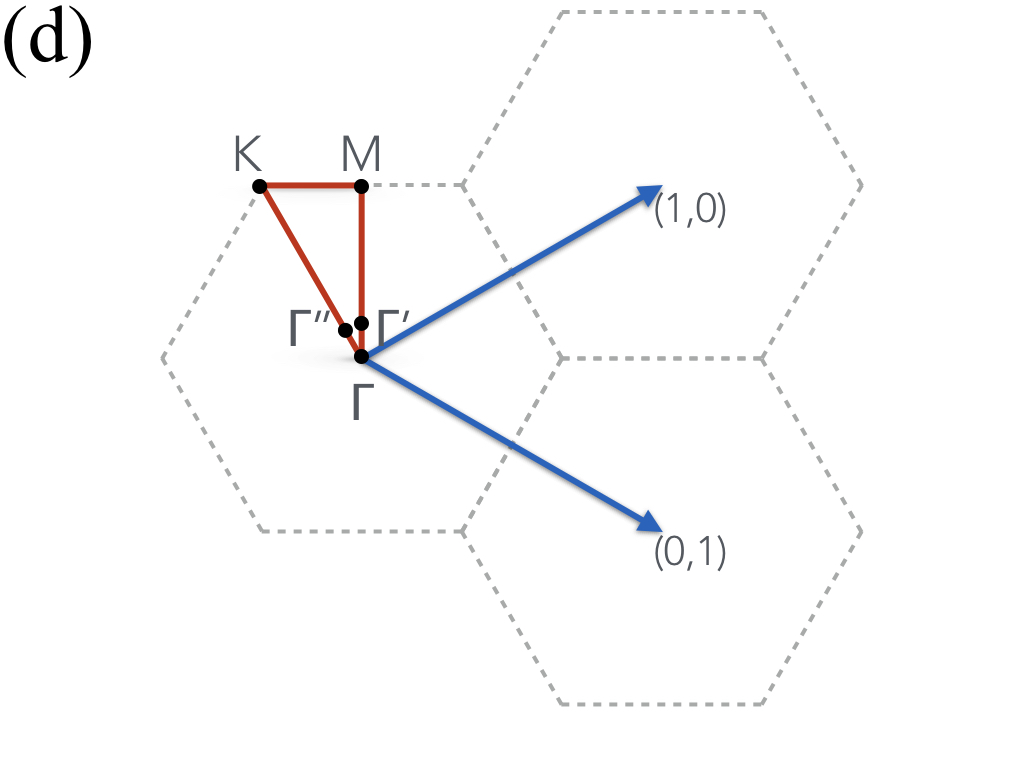}}
\caption{(Color online.) (a) The Yb triangular lattice with ${\bs a}_1, 
{\bs a}_2, {\bs a}_3$ bonds. (b) The spinon band structure for
$\Delta = 0.6B$ and ${t_2/t_1=0.2}$ (this value is optimized for 
the variational energy; see main text).
(c) A schematic illustration of the spinon band structure
and the particle-hole excitation for the zero momentum transfer.
(d) The Brillouin zone of the triangular lattice, with high-symmetry 
points and the basis vectors (in r.l.u. coordinates) highlighted.}
\label{fig1}
\end{figure}
%-----------------------------

\section{Coupling to the magnetic field}
\label{sec3}

Unlike the electron, the fermionic spinon is a charge neutral 
object and does not couple to the external magnetic field via the 
conventional Lorentz coupling. Here, we point out that the prior 
theory on the organic spin liquid material~\cite{kappaET} 
$\kappa$-(ET)$_2$Cu$_2$(CN)$_3$ has actually invoked the 
interesting Lorentz coupling of the spinons to the external 
magnetic field {\it indirectly} through the internal U(1) 
gauge flux~\cite{PhysRevB.73.155115}. This is because the 
organic material $\kappa$-(ET)$_2$Cu$_2$(CN)$_3$ is in the weak 
Mott regime where the charge gap is small and the four-spin ring 
exchange interaction can be significant~\cite{Motrunich2005}. 
It is the four-spin ring exchange that connects and transfers 
the external magnetic flux to the internal emergent U(1) gauge 
flux~\cite{PhysRevB.73.155115}. In contrast, the $4f$ electrons 
of the Yb ions are in the strong Mott regime and is very localized. 
As we have explained, the effective spin ${\boldsymbol S}_i$ arises 
from the strong spin-orbit coupling (SOC) and crystal electric field 
splitting, and the four-spin ring exchange is strongly suppressed 
due to the very large on-site interaction of the $4f$ electrons. 
Therefore, the orbital coupling to the magnetic field of the spinons 
in the organic spin liquid does not apply to YbMgGaO$_4$, and there 
will not be the spinon Landau level nor Hofstadter band structures.  
Although a strong magnetic field would polarize the Yb local moments 
along the field direction and thus destabilizes the spin liquid state, 
in the weak field regime, the field does not change the spin liquid 
ground state and the spinon remains to be a valid description of the 
magnetic excitation. From the above argument, if YbMgGaO$_4$ ground 
state is a spinon Fermi surface QSL, the appropriate spinon mean-field 
Hamiltonian for YbMgGaO$_4$ in a weak external magnetic field should be 
\begin{eqnarray}
H_{\text{MF}_{\text h}} &=& - t_1 \sum_{\langle ij \rangle,\alpha}
f^{\dagger}_{i\alpha} f^{}_{j\alpha}
- t_2 \sum_{\langle\langle ij \rangle\rangle,\alpha}
f^{\dagger}_{i\alpha} f^{}_{j\alpha}
\nonumber \\
&& 
- \mu \sum_{i,\alpha} f^{\dagger}_{i\alpha} f^{}_{i\alpha} 
- \sum_{i,\alpha\beta}
 g_z^{} \mu_{\text{B}}^{} h_z^{} f^{\dagger}_{i\alpha}
 \frac{\sigma^z_{\alpha\beta}}{2} f^{}_{i\beta},
\label{eq2}
\end{eqnarray}
where only Zeeman coupling is needed, and $g_z$ is the
Land$\acute{\text e}$ factors for the field normal to 
the triangular plane of the Yb atoms, respectively. 
The mean-field Hamiltonian in Eq.~\eqref{eq2} will be 
the basis of the analysis below.

\section{Spectral weight shifts from the free-spinon mean-field theory}
\label{sec4}

As the magnetic field is varied, the microscopic spin Hamiltonian that includes
the Zeeman coupling is modified. Thus, for each magnetic field, the spinon hopping 
and the chemical potential in Eq.~\eqref{eq2} need to be adjusted to optimize the 
variational energy of the microscopic spin Hamiltonian $H_{\text{Spin-h}}$ that is
\begin{eqnarray}
H_{\text{Spin-h}} &=& \sum_{\langle ij \rangle}
\big[ J_{zz}^{} S^z_i S^z_j + J_{\pm}^{} (S^+_i S^-_j + S^-_i S^+_j)
\nonumber \\
&& + J_{\pm\pm}
(\gamma_{ij}^{} S^+_i S^+_j  +\gamma_{ij}^{\ast} S^-_i S^-_j)
\nonumber \\
&& -\frac{{\mathbb i} J_{z\pm}}{2}
\big( ( \gamma^{\ast}_{ij} S^+_i - \gamma_{ij}^{} S^-_i ) S_j^z
+ \langle  i \leftrightarrow j \rangle
 \big) \big]
\nonumber \\
&& - \sum_i g_z^{} \mu_{\text{B}}^{} h_z^{} S_i^z .
\label{eq3}
\end{eqnarray}
Here $\gamma_{ij}$'s are the bond-dependent phase variables 
that arises from the spin-orbit coupling of the Yb $4f$ 
electrons~\cite{Yuesheng2015,YaodongPRB,Yaodong201608,Yaodong201612},
and ${\gamma_{ij} = 1, e^{{\mathbb i}\,2\pi/3}, e^{-{\mathbb i}\,2\pi/3}}$ for $ij$
along the ${\bs a}_1$, ${\bs a}_2$, ${\bs a}_3$ bond, respectively.
The recent polarized neutron scattering measurement has provided a 
clear support for the above nearest-neighbor anisotropic spin Hamiltonian
for YbMgGaO$_4$~\cite{Toth1705}. Moreover, the experiment did not 
find strong signature of the exchange disorder~\cite{Toth1705,yuesheng1702}. 
In the following calculation of this paper, 
we set ${J_{\pm} = 0.915 J_{zz}}$~\cite{Yuesheng2015,YaodongPRB}. 
The $z$-direction magnetic field shifts the chemical potential for 
the spin-$\uparrow$ and spin-$\downarrow$ spinons up and down such 
that the spinon excitations are given by
\begin{eqnarray}
\xi^{}_{\uparrow} ({\boldsymbol k})   &=& \epsilon ({\boldsymbol k})
- \mu^{}_{\uparrow} \equiv  \epsilon ({\boldsymbol k}) - (\mu^{}
+ \frac{g_z  \mu_{\text{B}} h_z}{2}),
\\
\xi^{}_{\downarrow} ({\boldsymbol k}) &=& \epsilon ({\boldsymbol k})
- \mu^{}_{\downarrow} \equiv  \epsilon ({\boldsymbol k})
- (\mu^{} - \frac{g_z  \mu_{\text{B}} h_z}{2}),
\end{eqnarray}
where $\epsilon ({\boldsymbol k})$ is the dispersion that is obtained
from the first line of Eq.~\eqref{eq2}. In Fig.~\ref{fig1}, we plot
the mean-field dispersions of the spinons in the magnetic field,     
where the spin up and spin down spinons have different Fermi surfaces.
Therefore, in the weak field regime, the system remains gapless.

\begin{figure}[t]
%\centering
\includegraphics[width=.23\textwidth]{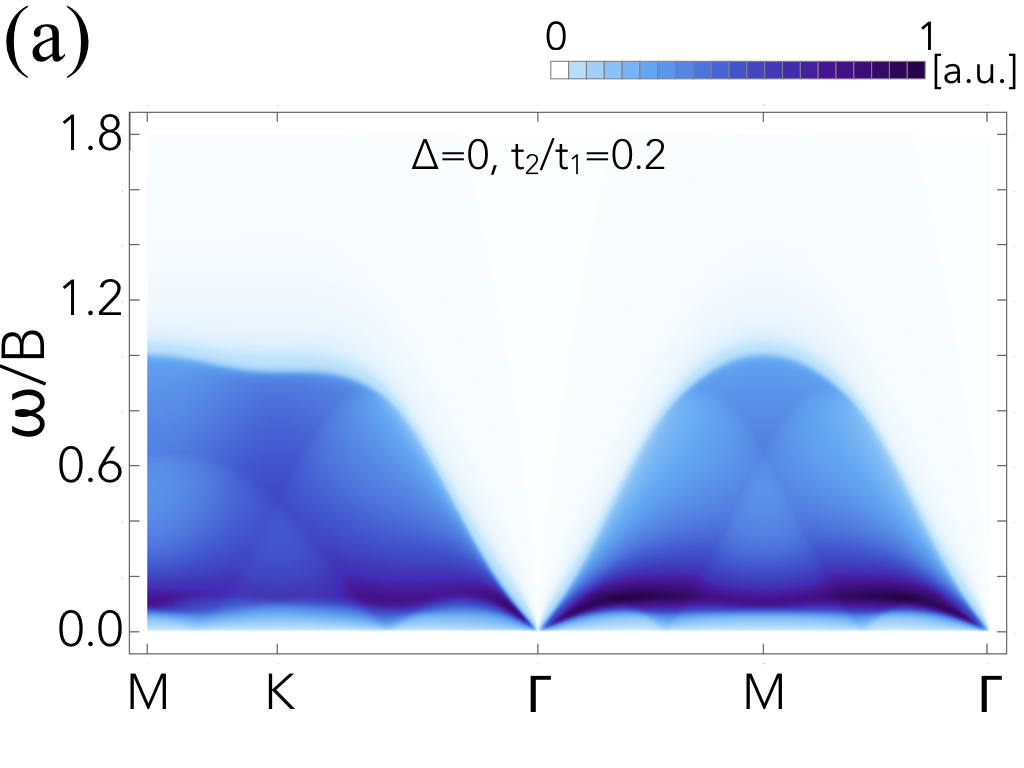}
\includegraphics[width=.23\textwidth]{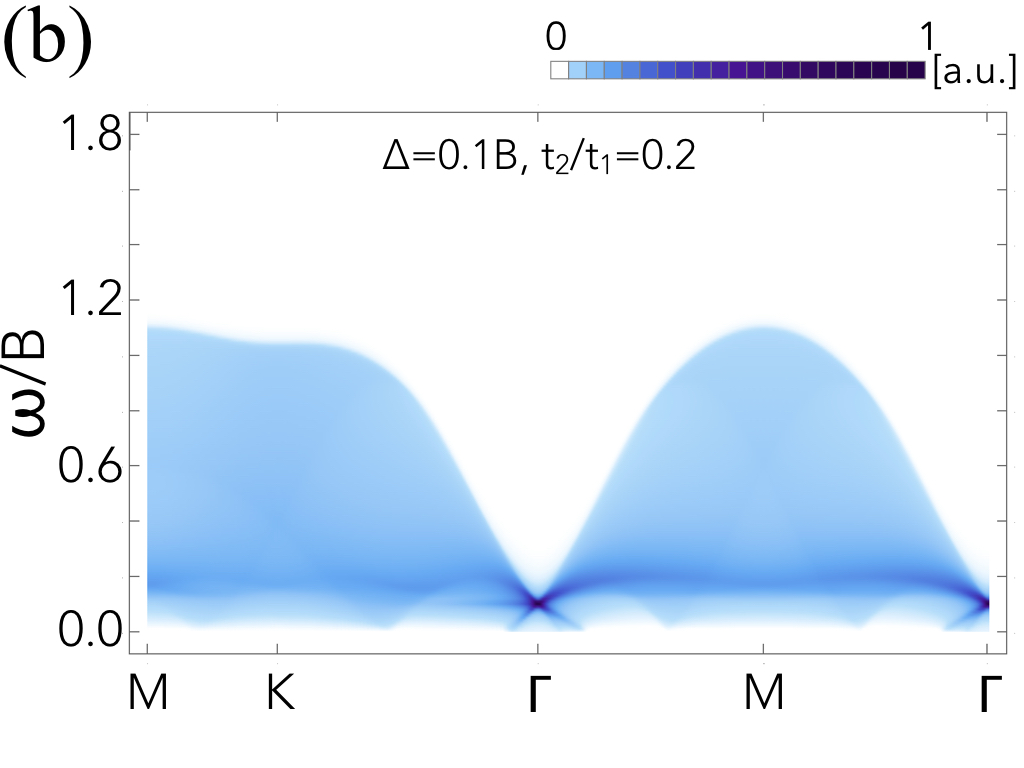}
\includegraphics[width=.23\textwidth]{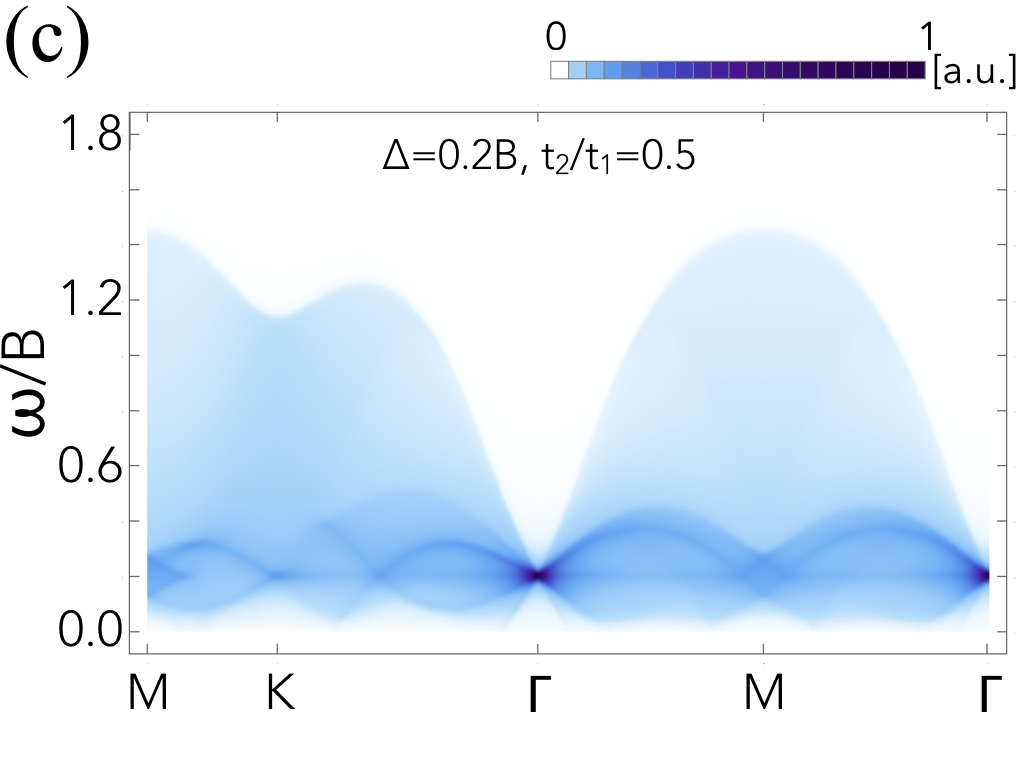}
\includegraphics[width=.23\textwidth]{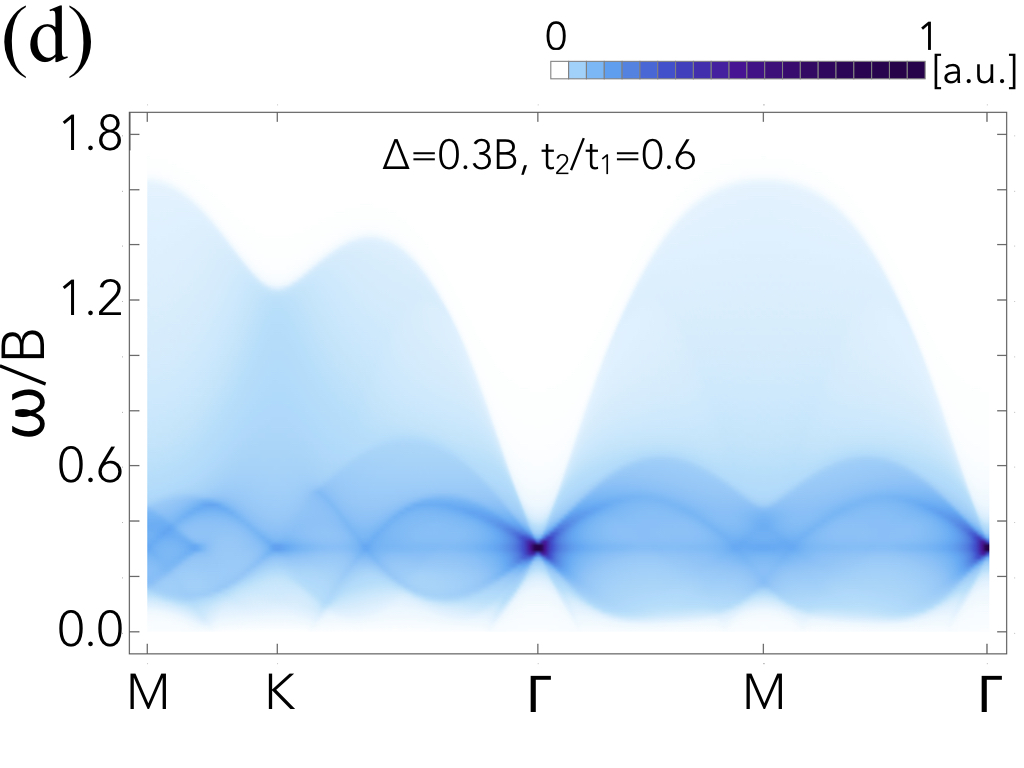}
\includegraphics[width=.23\textwidth]{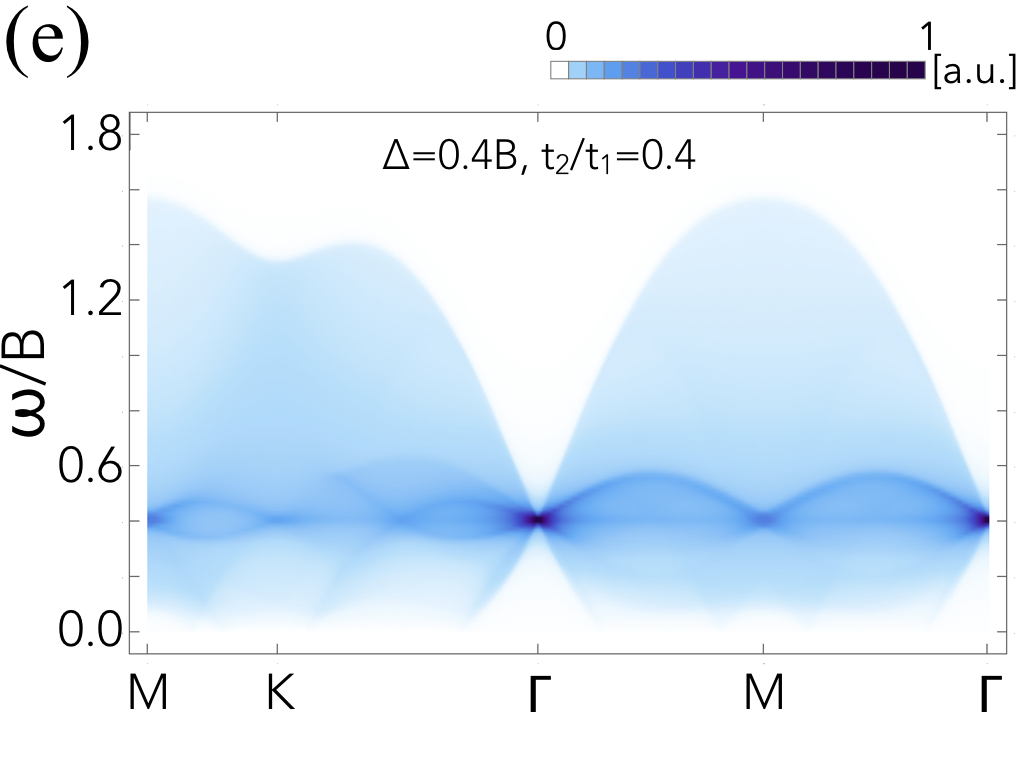}
\includegraphics[width=.23\textwidth]{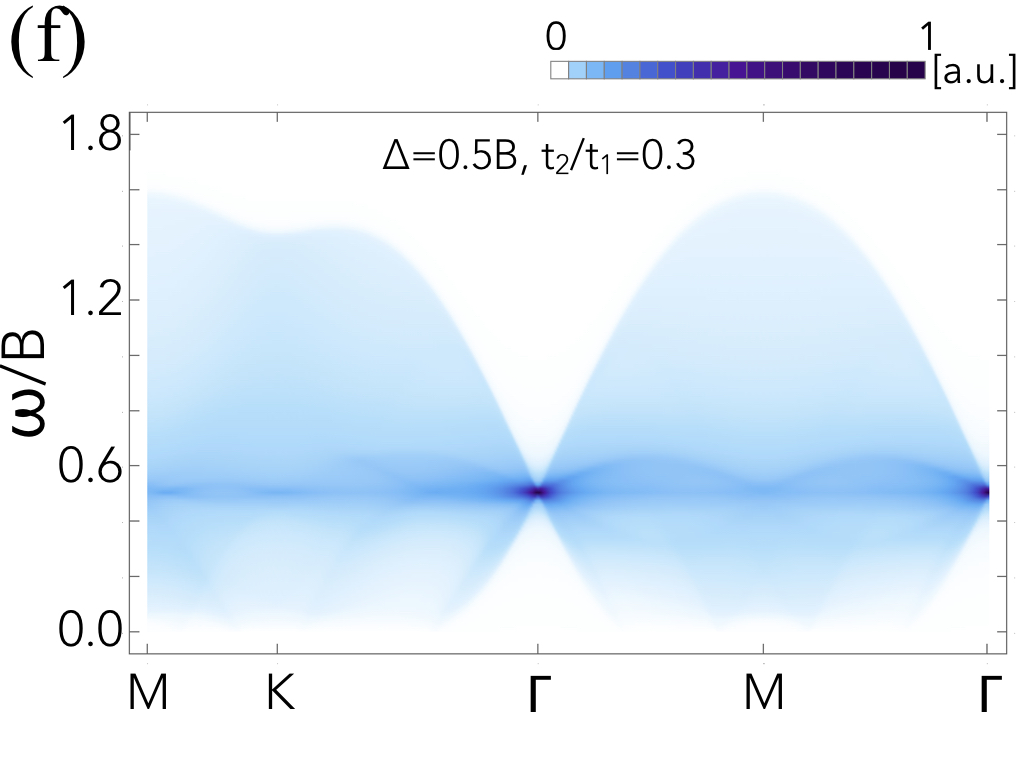}
\includegraphics[width=.23\textwidth]{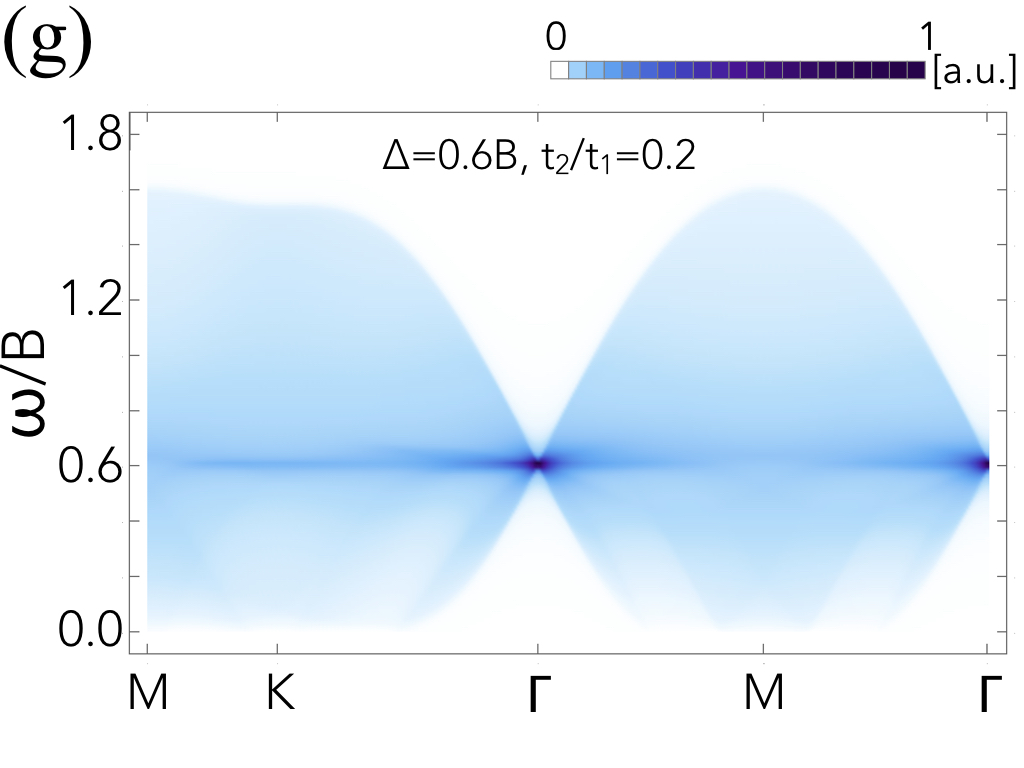}
\includegraphics[width=.23\textwidth]{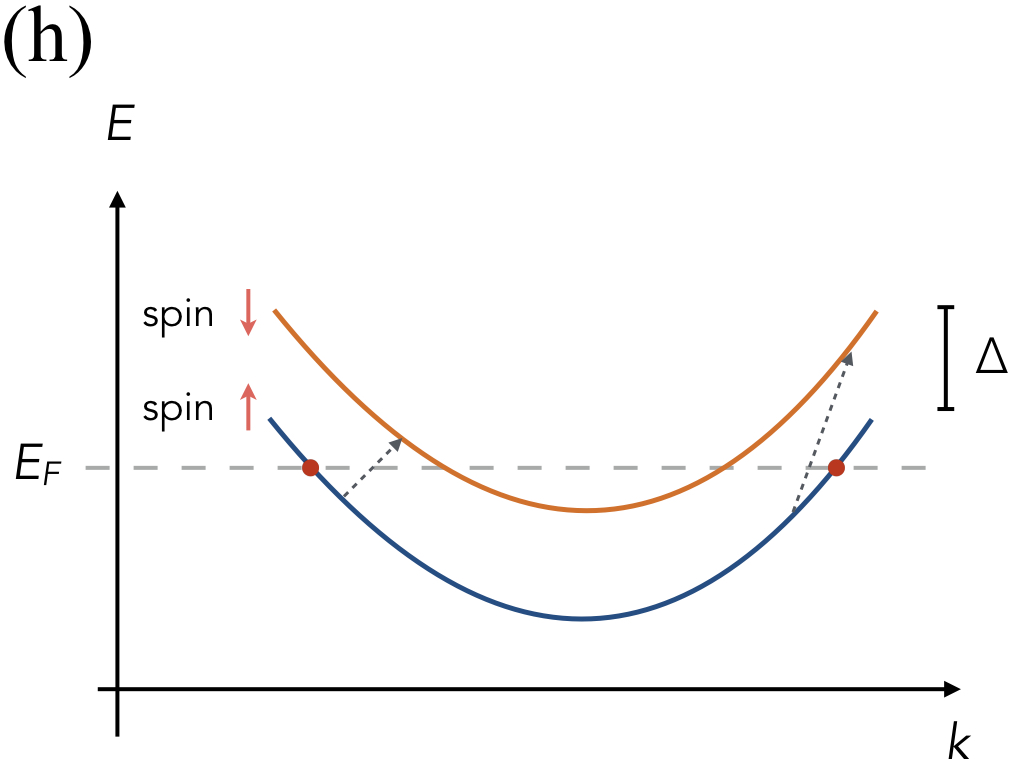}
\caption{(a-g) Dynamic spin structure factors for free spinon 
theory with $z$-direction magnetic field up to $0.6B$, 
where ${B=9.6t_1}$ is the bandwidth for the free spinon 
theory without the field in Eq.~\eqref{Hmf}.
The values of $t_2/t_1$ are optimized from the variational energy.
(h) Illustration of the particle-hole excitations with small momenta. 
Such excitations for each $\bs q$ are degenerate at zero field,  
and the $2$-fold degeneracy is lifted as soon as the field is turned on.
}
\label{fig2}
\end{figure}

In the inelastic neutron scattering measurement, the neutron spin flip
excites the spinon particle-hole pairs across the spinon Fermi surface.
The spin-flip event corresponds to the inter-band particle-hole excitation,
and we will focus on this process in the following discussion.
In the free-spinon mean-field theory, the energy and momentum change of 
the neutron, $\omega$ and ${\boldsymbol p}$, is shared by the one spinon 
particle-hole pair, and we have
\begin{eqnarray}
 {\boldsymbol p} &=& {\boldsymbol k}_1 - {\boldsymbol k}_2 , \\
 \omega ({\boldsymbol p}) &=& \xi_{\downarrow}^{}({\boldsymbol k}_1)
                            - \xi_{\uparrow}^{} ({\boldsymbol k}_2).
\end{eqnarray}
In the mean-field theory, the field essentially breaks the degenerate
spinon bands by separating the dispersions of spin-$\uparrow$ and 
spin-$\downarrow$ spinon bands in energy with a Zeeman splitting.
Thus, there exists a large density of particle-hole excitations 
at the zero momentum transfer with the energy ${\omega ({\boldsymbol 0}) 
= g_z \mu_{\text B}h_z \equiv \Delta}$ (see the illustration in 
Fig.~\ref{fig1}b). We thus expect a spectral peak at the 
$\Gamma$ point with a finite energy transfer $\Delta$. In the absence of 
the magnetic field, we have shown in Refs.~\onlinecite{YaoShenNature,Yaodong201612} 
that the spectral weight at the finite energy is suppressed at the 
$\Gamma$ point.

In Fig.~\ref{fig2}, we calculate and plot the dynamic spin structure 
factor ${\mathcal S}({\boldsymbol q},\omega)$ for various values of 
the magnetic field $h_z$ on a lattice of size ${200\times200}$. 
Here ${\mathcal S}({\boldsymbol q},\omega)$ selects
the spin-flipping events and is given by
\begin{eqnarray}
{\mathcal S}({\boldsymbol q},\omega) &=& \frac{1}{N}\sum_{i,j} 
e^{{\mathbb i} \, {\bs q} \cdot ({\bs r}_i - {\bs r}_j)} 
\int_t e^{-{\mathbb i} \, \omega t}
\langle S^-_{{\bs r}_i} (t) S^+_{{\bs r}_j} (0) \rangle, 
\end{eqnarray}
where $N$ is the system size and the expectation is taken with respect 
to the spinon Fermi surface ground state of the mean-field Hamiltonian. 
In the above equation, the time integration yields a delta function for the
energy conservation, and  we replace the delta function by $\delta (\omega)
= \frac{\eta/\pi}{ \omega^2 + \eta^2}$ with $\eta = 0.1t_1$ in 
the calculation. First we observe that the spectrum remains gapless 
at the weak field regime. Second, we indeed find that the 
spectral weight at the $\Gamma$ point occurs at the energy transfer 
$\Delta$ due to the splitted structure of the spinon bands that were 
discussed previously.

Apart from the enhancement of the spectral intensity at the $\Gamma$ 
point and the Zeeman splitting energy, there is an interesting spectral 
crossing near the Zeeman splitting energy $\Delta$ around the $\Gamma$ point.
This is the unqiue consequence of the external magnetic field on 
the spinon continuum. To understand this phenomenon, 
we start from the vertical particle-hole transition in Fig.~\ref{fig1}b and
slightly tilt the transition such that the momentum transfer
of the neutron is finite but small (see Fig.~\ref{fig2}h).
Depending on the tilted direction of the momentum, the energy
transfer can be greater or smaller than $\Delta$ and take
the value ${\Delta \pm  {\boldsymbol v} \cdot {\boldsymbol q} }$,
where we have linearized the dispersion and
${\boldsymbol v} \approx {\boldsymbol v}_{\text F}$
in the weak field limit. In principle, the velocity ${\boldsymbol v}$ 
would depend on the momenta of the spinon particle and spinon hole, but 
for the convenience of presentation, such dependence is not indicated. 
The neutron energy transfer is thus located within the energy range 
${({\Delta - v q}, {\Delta +  v q })}$, and this explains the upper and 
lower excitation edges near the $\Gamma$ point.

The above behaviors of spinon continuum in the weak magnetic 
field are qualitatively different from what one would expect for
the magnon-like excitation. For the magnons that are integer-spin 
excitations, the magnetic field directly couples to the magnon, 
most often shifts the whole magnon band by gapping out the low-energy modes.
The magnon lifetime becomes longer, the magnon quasi-particle become 
sharper, and the magnon band would be narrowed. In contrast, for
the spinons that are spin-1/2 excitations, the magnetic field 
shfits the spin-up spinon and the spin-down spinon bands 
oppositely, and the spin-up and spin-down spinons
are combined together to give the inter-band particle-hole contribution in
the dynamic spin structure factor.  
The spinon continuum in the magnetic field is 
sensitive to the dispersions of both spinon bands and thus 
reflects the fractionalized nature of the magnetic excitations. 
It is hard to imagine the broad excitation continuum,
the {\it spectral crossing} at $\Gamma$ and ${\omega=\Delta}$, 
and the {\it upper and lower excitation edges} near the $\Gamma$
point in Fig.~\ref{fig2} can be obtained from the magnon-like 
excitation under the magnetic field.

\begin{figure}[t]
\includegraphics[width=4.25cm]{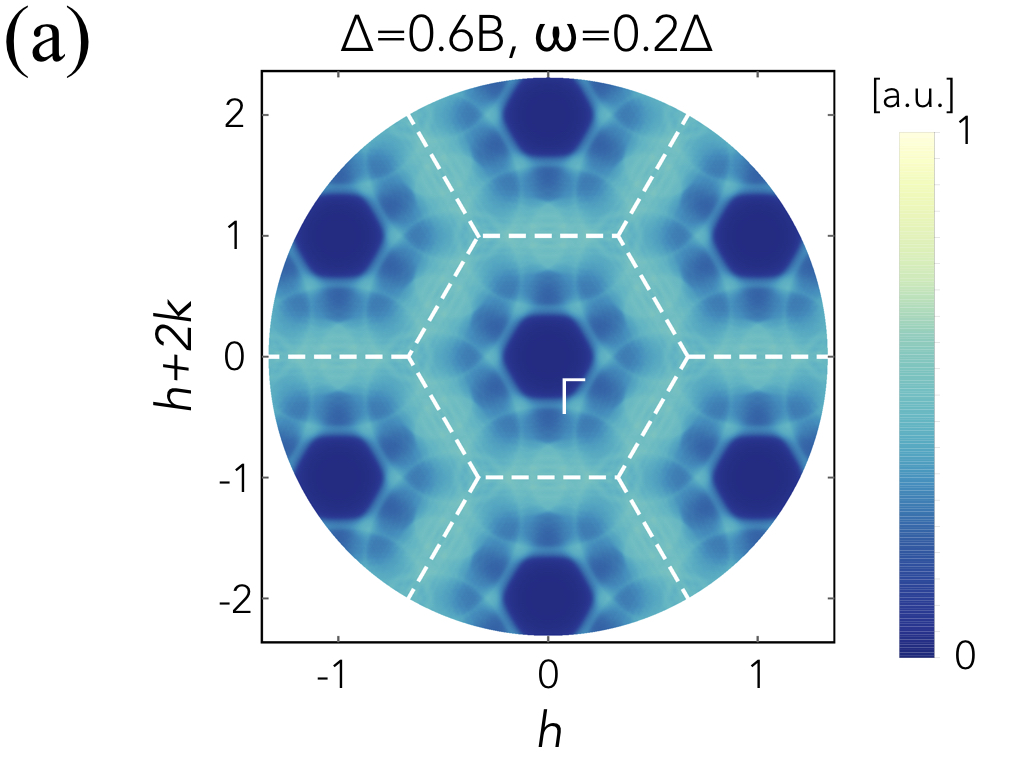}
\includegraphics[width=4.25cm]{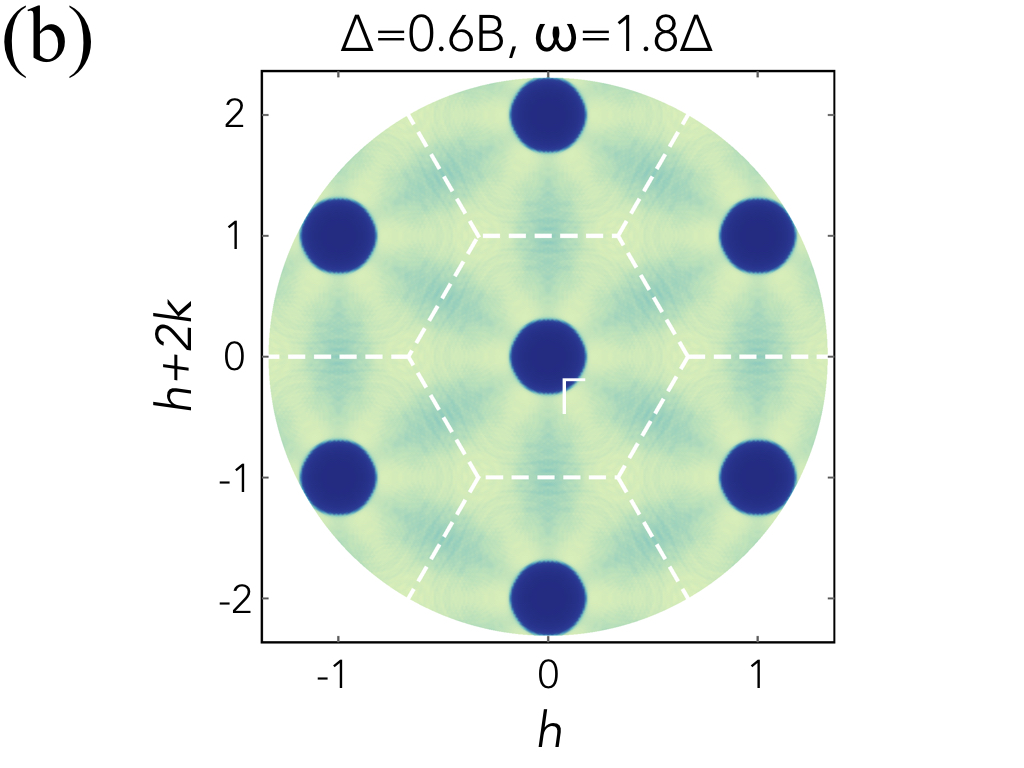}
\caption{Momentum resolved dynamic spin structure factor at
the energy cuts above and below the Zeeman splitting $\Delta$.
(a) and (b) share the same intensity bar. For both plots, the 
spectral weights in the region near the $\Gamma$ point are 
suppressed.
}
\label{fig3}
\end{figure}

Another feature of the spinon continuum in the weak magnetic field
is the suppression of the overall intensity. Originally at the zero
field, the spectral weight is suppressed above an upper excitation 
edge (see Fig.~\ref{fig2}a). The weak magnetic fields create spectral
weights near the $\Gamma$ point at finite energies, i.e. in the
regions where the spectral weights are suppressed at the zero field. 
Therefore, the overall intensity of the continuum is suppressed at
the small magnetic field.

To further manifest two excitation edges near the $\Gamma$ point
and ${\omega=\Delta}$, we depict the dynamic spin structure factor
in the Brillouin zone for different neutron energy transfers in 
Fig.~\ref{fig3}. The intensity distribution within the Brillouin 
zone further reflects the variation of the spinon band structure 
under the magnetic field. For the energy and the momentum below
the lower excitation edge in Fig.~\ref{fig2}, the spectral   
weight is strongly suppressed, leading to a reduced spectral 
intensity around the $\Gamma$ points (see Fig.~\ref{fig3}a). 
Likewise, for the energy above the upper excitation edge in Fig.~\ref{fig2}, 
the spectral intensity around the $\Gamma$ points is similarly
suppressed (see Fig.~\ref{fig3}b). Nevertheless, for energy right 
at $\Delta$, the spectral intensity around the $\Gamma$ points is 
not suppressed due to the reason that was explained previously.

\begin{figure}[t]
%\centering
\includegraphics[width=.23\textwidth]{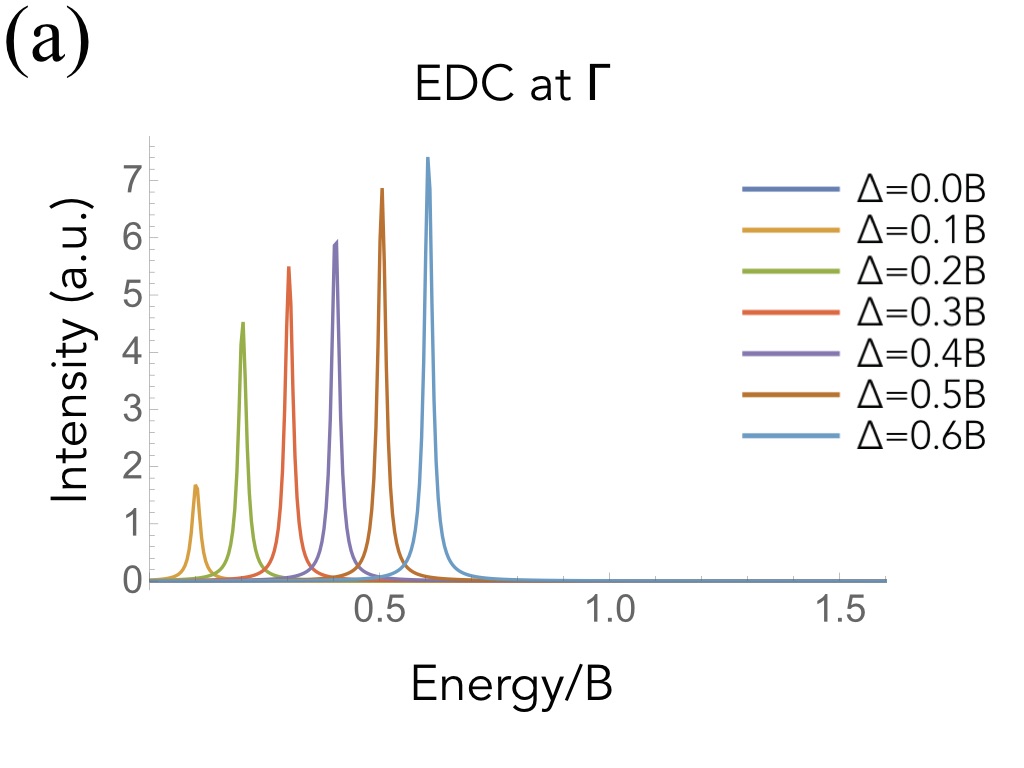}
\includegraphics[width=.23\textwidth]{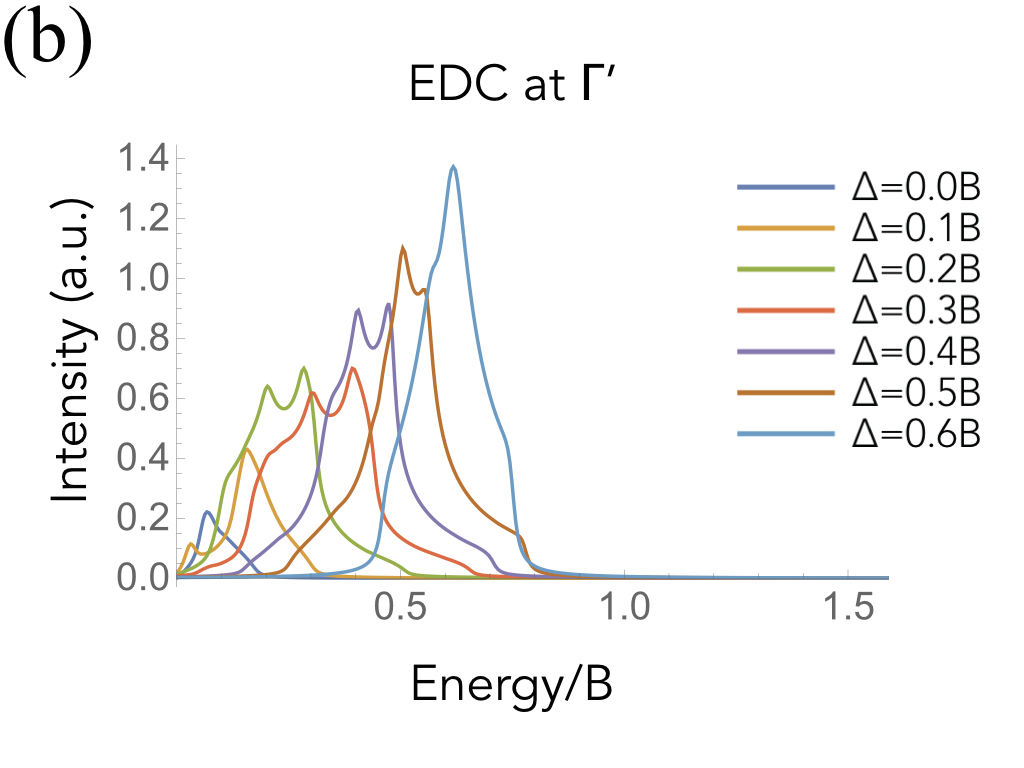}
\caption{ 
Energy-dependent curves of the dynamic spin structure factor 
at (a) $\Gamma$ and (b) $\Gamma^\p$ (see Fig.~\ref{fig1}d). 
Right at $\Gamma$, there is a narrow
Zeeman peak for nonzero fields whose position shifts 
with the field. Away from $\Gamma$, there is a broad 
continuum corresponding to the spinon particle-hole 
excitations. Note that the very low-energy part of spectral 
weight is underestimated in the mean-field theory due 
to the neglecting of the gauge fluctuation~\cite{Patrick1992}. 
}
\label{fig4}
\end{figure}

The energy distribution curve (EDC) of ${\mathcal S}({\boldsymbol q},\omega)$
at certain momenta is a common measurement with inelastic neutron     
scattering that could reveal important information of the ground state.
We depict the EDC in Fig.~\ref{fig4}. The peak at the $\Gamma$ point and
the shift of the peak position under the magnetic field are the most salient
experimental features, and can be readily probed using inelastic neutron
scattering and/or optical measurements. Off the $\Gamma$ point, the
spectral peak in the EDC becomes broad since the finite momentum
transfer allows a range of energies for the spinon particle-hole continuum,
and the energy range of the peak is bounded by the upper and lower 
excitation edges.

Finally, we comment on the caveat of the mean-field theory. 
In the mean-field theory for the spinons, we have ignored the 
spinon-gauge coupling. The gapless U(1) gauge photon is expected to 
play an important role at low energies. For example, the 
Yukawa coupling between the fermionic spinons and the gapless 
U(1) gauge photon would give rise a self-energy correction
to the spinon Green's function and thus enhance the low-energy
density of states~\cite{Patrick1992,LeeLeePRL}. 
Therefore, the inelastic neutron scattering process that excites 
the spinon particle-hole pair, would have an enhanced spectral 
weight at the low energies. This property is not captured in the 
spinon mean-field theory. We thus expect the very low energy 
spectral weights in Figs.~\ref{fig2},\ref{fig4} and also in 
Fig.~\ref{sfig1} to be enhanced when the gauge flucutation is 
included. Moreover, the slight enhancement of the overall bandwidth 
of the spinon continuum in the field is probably a mean-field artifact as
well because the spinon bandwidth is expected to be set 
by the exchange interaction of the system.

\section{The RPA correction from the anisotropic interaction}
\label{sec5}

As we have proposed in Ref.~\onlinecite{YaodongPRB}, the anisotropic 
spin exchange terms $J_{\pm\pm}$ and $J_{z\pm}$ from the strong SOC in 
Eq.~\eqref{eq3} is likely to play an important role in stabilizing the 
QSL ground state. The SOC is further suggested to be responsible for the weak 
spectral peak at the M point~\cite{YaoShenNature,Yaodong201608,Yaodong201612}.
Here we consider the effect of the anisotropic spin interaction 
on the dynamic spin structure factors following a phenomenological 
approach introduced in Refs.~\onlinecite{Yaodong201612,PhysRevLett.82.2915}. 
Starting from the free-spinon theory $H_{\text{MF}_{\text h}}$ and 
the corresponding susceptibility $\chi^0(\bs{q},\omega)$, we treat 
the anisotropic interaction $H_{\text{spin}}^\p$, that includes 
the $J_{\pm\pm}$ and $J_{z\pm}$ exchange interactions, as perturbations. 
The resulting magnetic susceptibility is calculated in the random 
phase approximation (RPA)~\cite{PhysRevLett.82.2915},
\begin{eqnarray}
\chi^\text{RPA} (\bs{q}, \omega) = 
[\bs{1} -\chi^0 (\bs{q}, \omega) {\cal J}(\bs{q})]^{-1} 
\chi^0 (\bs{q}, \omega),
\end{eqnarray}
where ${\cal J}(\bs{q})$ is the exchange matrix from $H_{\text{spin}}^\p$, 
\begin{eqnarray}
&&  {\cal J}({\bs q}) = \nonumber \\
&&    \begin{pmatrix}
    2 \left( u_{\bf q} - v_{\bf q} \right) J_{\pm\pm}
     & -2 \sqrt{3} w_{\bs q} J_{\pm\pm}
     & -\sqrt{3} w_{\bs q} J_{z\pm}  \\
     -2 \sqrt{3} w_{\bs q} J_{\pm\pm} &
     2 \left( v_{\bs q}   -u_{\bs q} \right) J_{\pm\pm}
     & \left(u_{\bs q} - v_{\bs q}  \right) J_{z\pm} \\
     -\sqrt{3} w_{\bs q} J_{z\pm}
     & \left(u_{\bs q} - v_{\bs q} \right) J_{z\pm}
     & %\left(u_{\bs q} + 2 v_{\bs q}\right) J_{zz} \\
     0
    \end{pmatrix} \quad \quad
\end{eqnarray}
with ${u_{\bs q} = \cos({\bs q}\cdot {\bs a}_1)}$, 
${v_{\bs q} = \frac{1}{2} [ \cos({\bs q} \cdot {\bs a}_2) 
+ \cos({\bs q}\cdot {\bs a}_3) ]}$, and 
${w_{\bs q} = \frac{1}{2} [ \cos({\bs q}\cdot {\bs a}_2) 
- \cos({\bs q}\cdot {\bs a}_3) ]}$.

\begin{figure}[t]
%\centering
{\includegraphics[width=.23\textwidth]{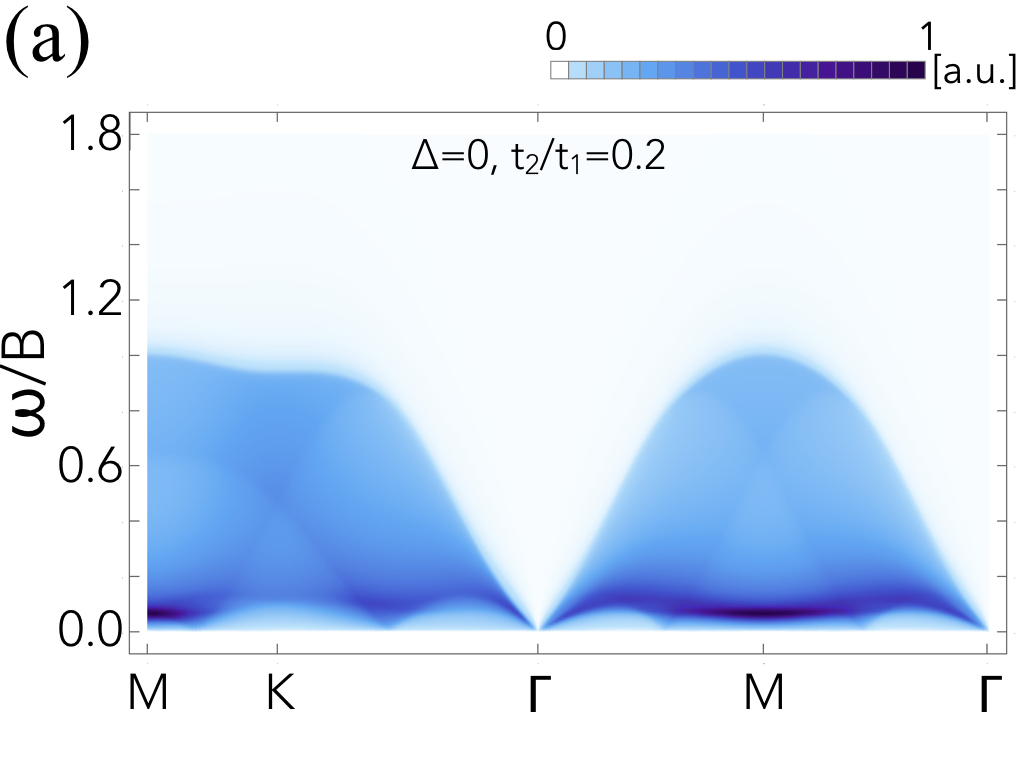}
\includegraphics[width=.23\textwidth]{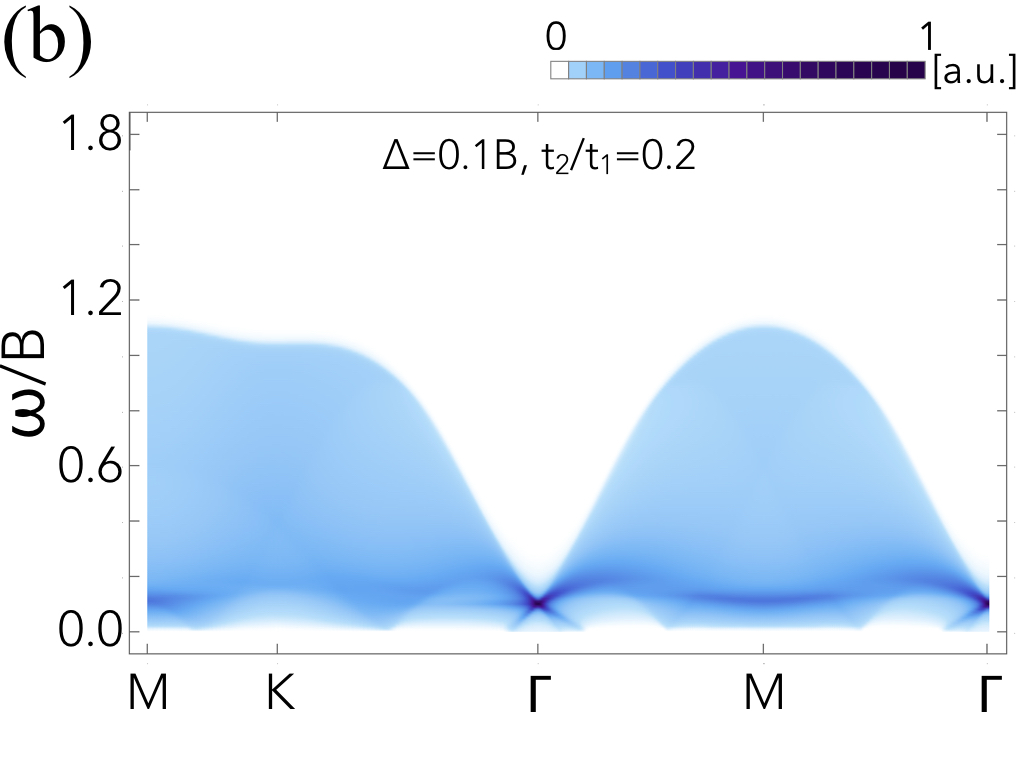}
\includegraphics[width=.23\textwidth]{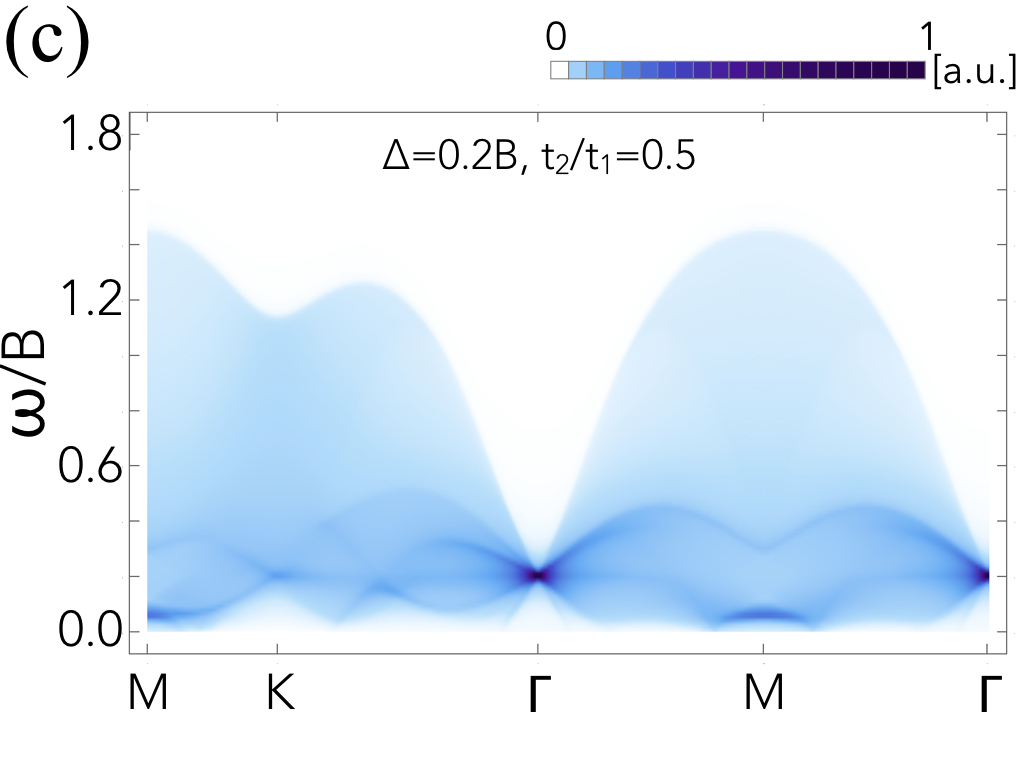}
\includegraphics[width=.23\textwidth]{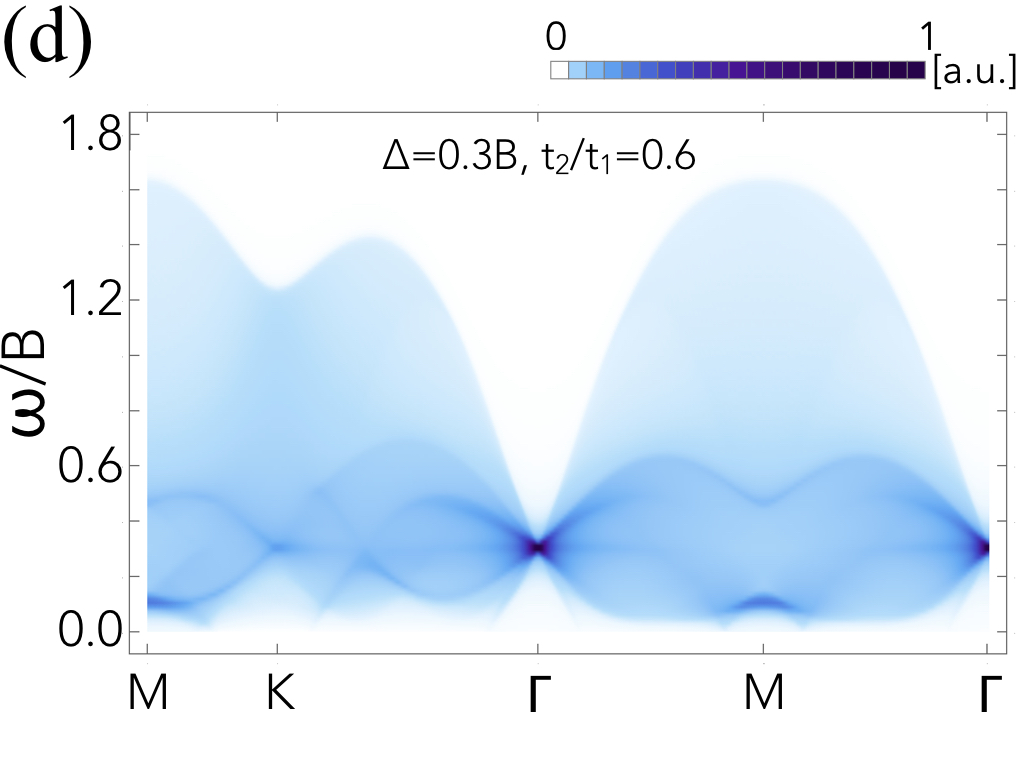}
\includegraphics[width=.23\textwidth]{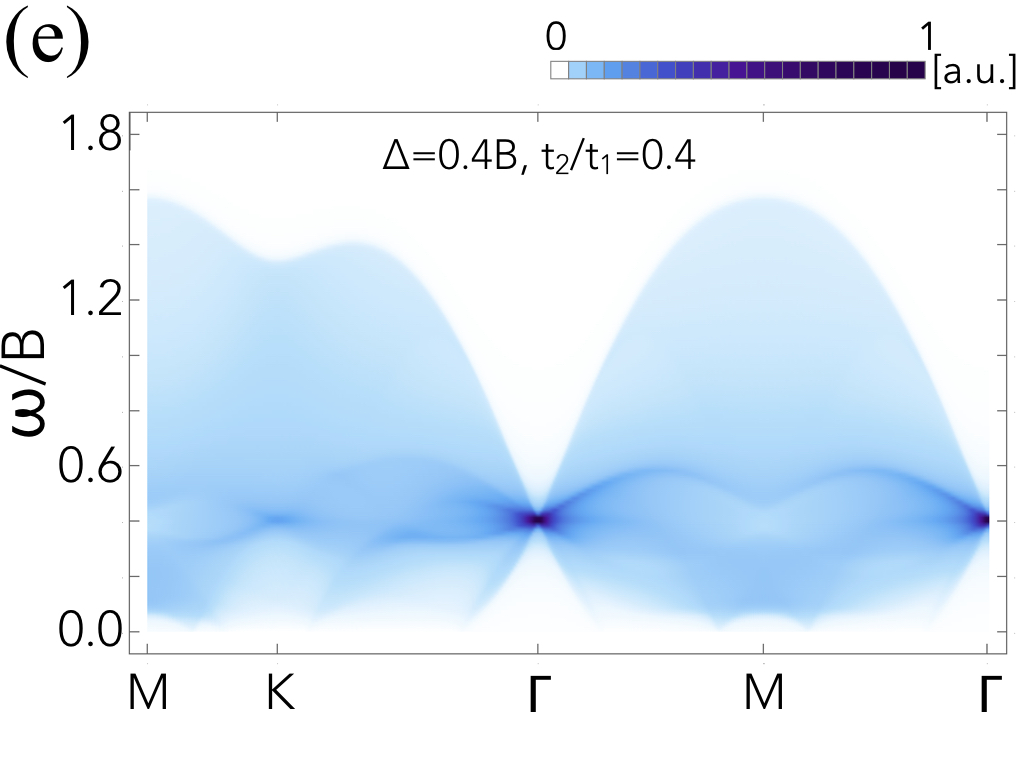}
\includegraphics[width=.23\textwidth]{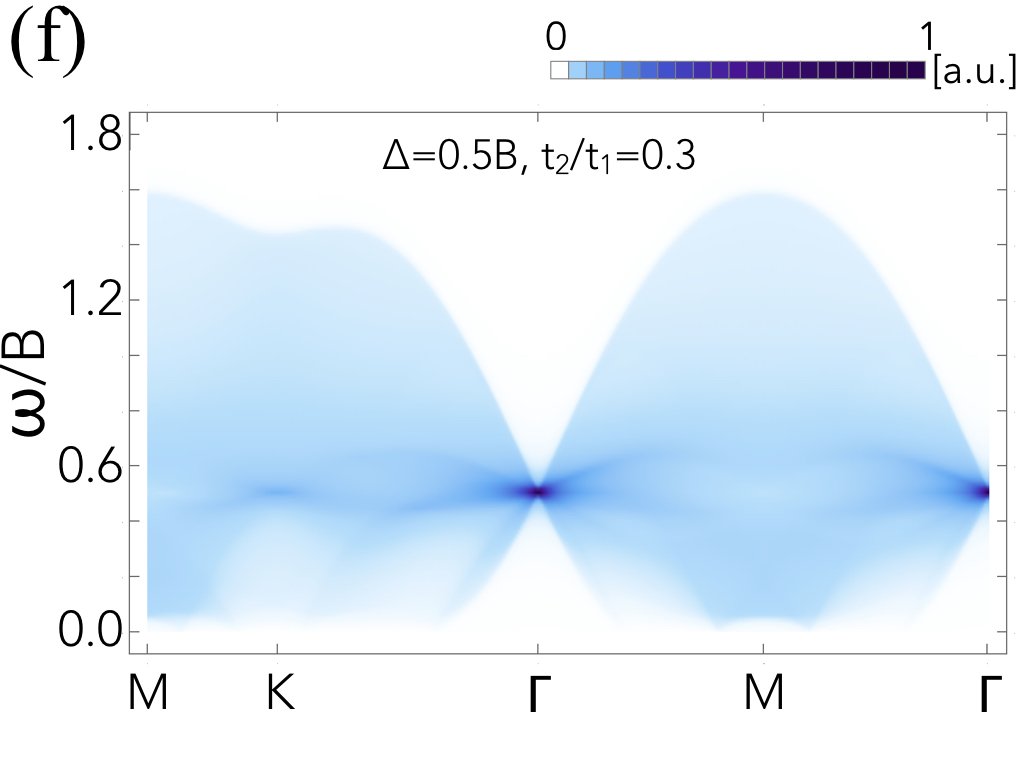}
\includegraphics[width=.23\textwidth]{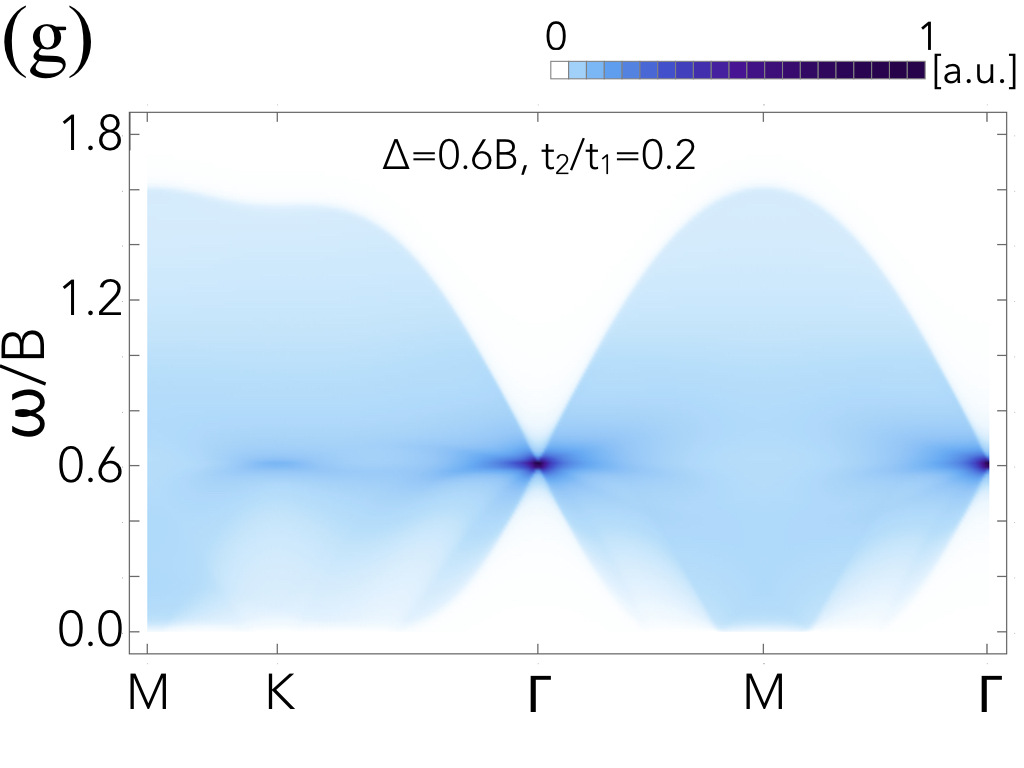}
\includegraphics[width=.23\textwidth]{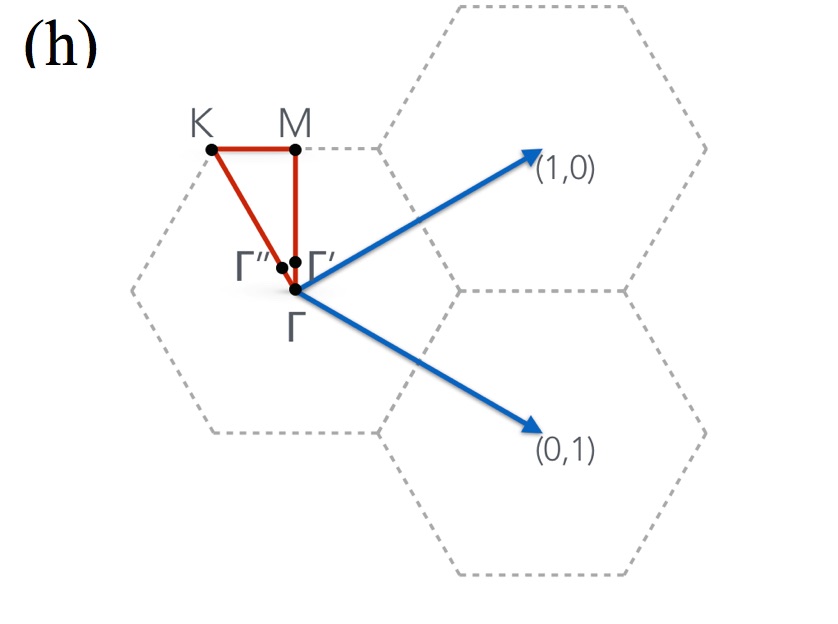}
}
 \caption{Dynamic spin structure factors for the interacting 
 spinon theory with external magnetic field along $z$-direction 
 up to $0.6B$, where the interaction is given by $H_{\text{Spin}}^\p$.
}
\label{sfig1}
\end{figure}

The RPA corrected dynamic spin structure factor is related 
$\chi^\text{RPA} (\bs{q}, \omega)$ by the equation 
${\cal S}^\text{RPA} (\bs{q}, \omega) = 
-\frac{1}{\pi} \text{Im} [\chi^\text{RPA} (\bs{q}, \omega)]^{+-}$.
The renormalized dynamic spin structure factor 
${\cal S}^\text{RPA} (\bs{q}, \omega)$ is shown 
in the Fig.~\ref{sfig1}, where we choose the 
parameters to be $J_{z\pm}/t_1 = 0.2$, $J_{\pm\pm}/t_1 = 0.35$.
From the results we conclude that the anisotropic exchange terms 
merely redistribute the spectral weight within the Brillouin zone 
and leave the qualitative features in the vicinity of 
the $\Gamma$ point mentioned in previous sections unaffected.

\section{Discussion}
\label{sec6}

Our prediction of the behaviors of the spinon continuum in the magnetic 
field relies on the spinon mean-field theory and the proposed spinon 
Fermi surface QSL state. In the calculation, we have applied the field 
along the $z$ direction ({\sl i.e.} normal to the triangular plane). 
Due to the spin rotation symmetry of the mean-field spinon Hamiltonian
in Eq.~\eqref{eq2}, we expect the orientation of the magnetic field 
to lead to qualitatively similar effects on the inelastic neutron spectrum. 
The microscopic spin model did not play a 
significant role in our experimental prediction, but the feasibility 
of the experiments in the magnetic field strongly relies on the fact 
that the microscopic interactions between the local moments are of 
the order of 1$\sim$10 Kelvins~\cite{Yuesheng2015}. 
A couple Tesla magnetic field applied to the material could readily 
lead to a visible effect on the spinon continuum. 
In contrast, the exchange couplings of other spin liquid candidate 
materials such as the triangular lattice 
organic spin liquids~\cite{kappaET,organics2,dmit} and 
herbertsmithite~\cite{YoungLee,mendels2007,PhysRevLett.98.107204} 
are of the order of 100K, and a much larger field is required there. 

The previous works on YbMgGaO$_4$~\cite{Martin2016,yuesheng1702} 
(including two of our earlier theoretical work~\cite{YaodongPRB,Yaodong201608}) 
have explored the strong field regime, where the magnetic excitations
are simply the {\it gapped} magnons with respect to an almost fully 
polarized state along the field directions. These works may provide 
useful information about the spin interaction, but do not give information 
about the spin quantum number fractionalization since the ground 
state is no longer a spin liquid state. 
In the zero field case~\cite{YaoShenNature,Yaodong201612} where 
the QSL physics was proposed~\cite{srep,Yuesheng2015}, 
we have shown that the spinon continuum 
develops a broad continuum in the momentum and energy domain.
In particular, due to the particle-hole excitation near the Fermi surface, 
there exists a ``V-shaped'' excitation edge at low energies 
near the $\Gamma$ point (see Fig.~\ref{fig2}a). 

In the {\it weak magnetic field} regime, the spinon bands are splitted. 
The spin-flipping particle-hole process is no longer degenerate 
with the spin-preserving particle-hole process. The inelastic 
neutron scattering measurement, that detects the  
correlations of the spin components perpendicular to the momentum,
includes both the spin-flipping and the spin-preserving processes. 
The spin-preserving process corresponds to the intra-band process,
and we would expect a ``V-shaped'' excitation edge at low energies 
near the $\Gamma$ point that is analogous to the zero-field results. 
In contrast, the spin-flipping process corresponds to the inter-band 
particle-hole excitation and is studied in this paper. For the inter-band
process, our theoretical predictions of\vspace{0.2cm} 
\\1) the spectral weight shifts, 
\\2) the spectral crossing at the $\Gamma$ point at the Zeeman splitting 
energy, 
\\ 3) the upper and lower excitation edges near the $\Gamma$ point, \vspace{0.1cm} 
\\
directly reveal the fractionalized spinon excitations. 
If these predictions for the weak field regime are confirmed experimentally, 
it will provide further support for the spinon Fermi surface QSL ground 
state in YbMgGaO$_4$.  

Apart from the spin quantum number fractionalization, the emergence of 
the Fermi statistics for the spinons is a rather unusual phenomenon. 
Although the particular spectral structure of the spinon continuum 
is a direct consequence of the spinon Fermi surface and the spinon 
Fermi statistics, directly confirming the
Fermi statistics is certainly desirable. The temperature dependence of the
dynamic spin structure factor could provide hints for the Fermi statistics.
Moreover, the spin-orbital-entangled nature of the Yb local moments may
provide a route to visualize the spinon Fermi surface. Clearly, the 
orbital degrees of freedom are sensistive to the ion position. 
The Yb local moment, that results from the spin-orbital entanglement,
may be more susceptible to the lattice degrees of freedom than
the conventional spin-only moment. Therefore, like the electron-phonon coupling
in Fermi liquids, one may expect a similar ``$2k_{\text F}$'' Kohn anomaly~\cite{senthil}
in the phonon spectrum that arises from the spinon-phonon coupling 
in YbMgGaO$_4$ and use the Kohn anomaly to construct the spinon Fermi surface.

Recently, there is an experimental work~\cite{Yuesheng1704} 
that proposes a scenario of nearest-neighbor resonant valence 
bond state for YbMgGaO$_4$ and claims ``the excitation continuum 
bears no obvious relation to spinons''. In fact, the nearest-neighbor 
resonant valence bond state on a frustrated lattice like the triangular 
lattice is precisely a $\mathbb{Z}_2$ spin liquid with gapped spinons and visons~\cite{Moessner2001,MoessnerSondhi,BalentsFisherGirvin}. 
Thus, the magnetic excitation revealed by the inelastic neutron scattering 
for such a $\mathbb{Z}_2$ spin liquid scenario has to be spinons. 
If such a senario is correct, doping this material will lead to 
superconductivity since the doped resonant valence bond state was 
suggested to one origin for the high temprature superconductivity 
in cuprates~\cite{Anderson1987,RevModPhys.78.17}
However, it is hard to reconcile the gapped spinon continuum 
with the particular dispersive continuum along high symmetry momenta 
in Ref.~\onlinecite{YaoShenNature}. 
Moreover, such a scenario does not 
seem to be compatible with the heat capacity, magnetic susceptibility 
and $\mu$SR results on this material~\cite{srep,YueshengmuSR,Martin2016}.  
On the theoretical side, the scenario of disorder was recently proposed~\cite{1703}. 
It might be an interesting possibility. The spectroscopic property 
about the magnetic excitation from this disorder scenario has not 
yet been given. The direct comparison and prediction with the 
inelastic neutron scattering results would certainly be desirable. 

While the main focus of this paper is on YbMgGaO$_4$, our results for the spinon
continuum may generally be applicable to other candidate materials with spinon Fermi surfaces. 
For the 6H-B phase of Ba$_3$NiSb$_2$O$_9$, the Curie-Weiss temperature is about $-75$K and sets
the bandwidth of the spinons~\cite{PhysRevLett.107.197204}. 
The spin-1 nature of the local moment enhances the Zeeman
coupling, so a $\sim$10T magnetic field would have a visible effect 
on the spinon dispersion. This may be useful to differentiate other 
theoretical proposals on this 
material~\cite{PhysRevLett.109.016402,PhysRevB.86.224409,PhysRevLett.108.087204}. 
If the ground state of this material is a spinon Fermi surface QSL, we expect 
similar effect of the spectral weight shift and spectral crossing may occur.

\section{Acknowledgements}

We acknowledge useful discussion with Jun Zhao and Patrick Lee, 
and constructive comments from an anonymous referee. 
This work is supported by the Ministry of Science 
and Technology of China with the Grant No.2016YFA0301001, 
the Start-Up Funds and the Program of First-Class 
University Construction of Fudan University, and the 
Thousand-Youth-Talent Program of China. G.C. thanks 
the hospitality of Prof Ying Ran at Boston College 
during the visit in January 2017 when this work is 
finalized.

\bibliography{Ref_Dec}

\end{document}